\documentclass[11pt]{article}
\pdfoutput=1

\usepackage{aas_macros,amsmath,amssymb,comment,cite,esint,graphicx,mathtools}
\usepackage[margin=.8in,letterpaper]{geometry}
\usepackage[colorlinks=true]{hyperref}
\usepackage[affil-it]{authblk}
\usepackage{url}
\usepackage{subcaption}
\usepackage{graphicx}

\usepackage{epsf,amsmath,bbold,amsfonts,stmaryrd}

\usepackage[utf8]{inputenc}
\usepackage{mathrsfs}
\usepackage{appendix}
\usepackage{amssymb}
\usepackage{float}
\usepackage{color}
\usepackage{cite}
\usepackage{hyperref}
\hypersetup{pageanchor=false}
\usepackage{indentfirst}
\usepackage{url}
\usepackage{float}
\usepackage{caption}
\usepackage[numbers,square,comma,sort&compress,merge]{natbib}
\usepackage{esint}
\usepackage{overpic}

\numberwithin{equation}{section}
\let\originalleft\left
\let\originalright\right
\renewcommand{\left}{\mathopen{}\mathclose\bgroup\originalleft}
\renewcommand{\right}{\aftergroup\egroup\originalright}
\mathcode`\*="8000
{\catcode`\*=\active\gdef*{\mathclose{}\,\mathopen{}}}

\newcommand{\ret}{\text{ret}}

\newcommand{\be}{\begin{equation}}
\newcommand{\ee}{\end{equation}}
\newcommand{\bea}{\setlength\arraycolsep{2pt} \begin{eqnarray}}
\newcommand{\eea}{\end{eqnarray}}
\newcommand{\nn}{\nonumber}
\newcommand{\mo}{\mathcal{O}}

\newcommand{\Len}{L_\text{en}}
\newcommand{\mi}{\text{min}}
\newcommand{\ma}{\text{max}}

\def \nn {\nonumber}

\begin{document}
\title{Polarized Images of Synchrotron Radiations  in Curved Spacetime}

\author{
Zezhou Hu$^{1}$, Yehui Hou$^{1}$, Haopeng Yan$^{2}$, Minyong Guo$^{3\ast}$
Bin Chen$^{1,2,4}$}
\date{}

\maketitle

\vspace{-10mm}

\begin{center}
{\it
$^1$Department of Physics, Peking University, No.5 Yiheyuan Rd, Beijing
100871, P.R. China\\\vspace{4mm}

$^2$Center for High Energy Physics, Peking University,
No.5 Yiheyuan Rd, Beijing 100871, P. R. China\\\vspace{4mm}

$^3$ Department of Physics, Beijing Normal University,
Beijing 100875, P. R. China\\\vspace{4mm}

$^4$ Collaborative Innovation Center of Quantum Matter,
No.5 Yiheyuan Rd, Beijing 100871, P. R. China\\\vspace{2mm}
}
\end{center}

\vspace{8mm}

\begin{abstract}
{In this work, we derive two formulas encoding the polarization direction and luminosity of synchrotron radiations from the moving electrons in curved spacetime under the geometric optics approximation. As an application, we further study the polarized images of synchrotron radiations from electron sources in Schwarzschild black hole spacetime with a vertical and uniform magnetic field. In particular, by focusing on the circular orbits of electrons on the equatorial plane, we show the polarized images of the synchrotron radiations from these orbits for different observational angles and discuss the variations of the polarization directions concerning the angles.}
\end{abstract}

\vfill{\footnotesize $\ast$ Corresponding author: minyongguo@bnu.edu.cn}

\maketitle

\newpage
\baselineskip 18pt
\section{Introduction}\label{sec1}
In 2019, the Event Horizon Telescope (EHT) Collaborations published the first images of a supermassive black hole in M87 \cite{EventHorizonTelescope:2019dse}. This opened a new window to study various problems in strong-field gravity and the physics of accretion disks via electromagnetic signatures. Very recently, the EHT Collaborations released the polarized images of the black hole\cite{EventHorizonTelescope:2021bee, EventHorizonTelescope:2021srq}, which reveal a bright ring of emission with a twisting polarization pattern and a prominently rotationally symmetric mode.

Over the past few decades, people have been trying to simulate the polarimetric images of black holes to understand the properties of the accretion disk and spacetime geometry\cite{Connors1980, Bromley:2001er, Broderick:2005jj, Li:2008zr, Schnittman:2009pm, Shcherbakov:2010ki, Gold:2016hld, Moscibrodzka:2017gdx, Jimenez-Rosales:2018mpc, Marin:2017rlw, Palumbo:2020flt}. For some recent studies on the polarized images of the black hole, see \cite{Lupsasca:2018tpp, EventHorizonTelescope:2021btj, Ricarte:2021frd, Gelles:2021kti, Zhang:2021hit, Karas:2021ltz, Qin:2021xvx, Lee:2022rtg}. In these works, the synchrotron radiation sources are considered charged fluids. The polarization vector is first defined in the locally flat comoving reference frame of fluids and then transformed to the laboratory frame with a non-flat metric. One way to determine the polarization vector radiated from the charged fluids is based on the Boltzmann equation\cite{10x} and employing numerical simulations. The other way is just to set the polarization vector perpendicular to both the wave vector of the radiation and the magnetic field in the fluid frame as a toy model \cite{EventHorizonTelescope:2021btj}. The former results are more realistic. However, the cost of numerical simulations is relatively high, and it is challenging to see how to decouple astrophysical and relativistic effects. Although the latter is a simplified model, it can help us intuitively understand the polarized image.

The polarized images originate from the polarized synchrotron radiations from electrons orbiting the black hole. Thus, theoretically, it is interesting to have a good insight into the polarized synchrotron radiations from electrons as spots orbiting a black hole. In this mechanism, the polarization vector can be extracted from the electromagnetic radiations of charged particles governed by the Maxwell equations. As for the electromagnetic radiations of charged particles in curved spacetime, a seminal work was made by DeWitt and Brehme (DB) in \cite{1960Radiation}, where they first studied the electromagnetic radiation damping in a gravitational field and gave the expressions of the electromagnetic Green's function and electromagnetic field tensor in a curved spacetime using the Hadamard expansion techniques. In particular, they found the electromagnetic Green's function would give rise to a "tail" term that invalidates Einstein's Equivalence Principle (EEP).
Moreover, the equations of motion of charged particles that they derived had a computation error, which was corrected by Hobbs in \cite{Hobbs:1968}. This problem was generalized to a charged body coupled to a Maxwell field on an arbitrary, fixed curved background by Quinn and Wald (QW) in \cite{Quinn:1996am}. In the electromagnetic case, they postulated a "comparison axiom" and then obtained their result, which exactly agreed with that of DeWitt and Brehme, as corrected by Hobbs.

In this work, we would like to consider electrons as source spots and investigate the polarized images of synchrotron radiations from electrons orbiting a black hole immersed in a uniform magnetic field. Based on DB's pioneering work \cite{1960Radiation}, we first work out two formulas encoding the direction and luminosity of the radiation in a covariant way. These formulas are valid for any curved spacetime. Then we focus on the circular orbits of electrons accelerated only by a uniform and vertical magnetic field on the equatorial plane in the Schwarzschild black hole spacetime and apply our formulas to study the polarized images of synchrotron radiations from these orbits. It is worth emphasizing that our formulas are wider than circular orbits. Compared to the method developed in \cite{EventHorizonTelescope:2021btj} where they consider the radiations of ionized fluids, the polarizations of radiations from collisionless charged particles we consider in this work are not only perpendicular to the magnetic field in the local rest frame of charged particles; instead, all the components of the radiations are included. In addition, we haven't introduced the degree of linear polarization in the current work. Though the fluid method applies to high-density plasma, it needs to introduce parameters of the accretion flow, such as the magnetic field strength, in the local rest frame, which further depends on the phenomenological assumption. The charged particles in low-density accretion disks may have a relatively long free path and be considered collisionless \cite{Zajacek:2018ycb}. Thus, our method developed in this paper has the potential to work in the magnetosphere with dilute plasma, where particle collisions can be neglected for dynamical processes such as jets and so on \cite{Rueda:2022fgz, Zahrani:2022fdd}. Moreover, since our formulas are all covariant, analyzing the radiations in this scheme in terms of a given background and an electromagnetic field has theoretical advantages.

The remaining parts of this paper are organized as follows. In Sec. \ref{sec2}, we work out the expressions of the luminosity and the polarization vector of synchrotron radiation after reviewing DB's work concisely. In Sec. \ref{sec3}, we complete other theoretical preparations for calculating the polarized images of synchrotron radiations around a Schwarzschild black hole. In Sec. \ref{sec4}, we show the polarized images for some specific examples. We summarize and conclude this work in Sec. \ref{sec5}.

\section{Synchrotron radiation in curved spacetime}\label{sec2}

In this section, we try to understand the synchrotron radiation of charged particles in a curved spacetime under geometric optics approximation. The electromagnetic radiations of moving charged particles have been discussed in \cite{1960Radiation, Hobbs:1968, Quinn:1996am}. Here we first give a brief review of DB's work, then we derive two novel formulas on the luminosity and polarization vector of radiations.

%We set up our research problems and models in this section. We focus our attention on a black hole immersed in uniform magnetic fields. Since we would like to consider polarizations of synchrotron radiations emitted from the relativistic charged particles moving near the black hole, first of all, we need to know the polarization vector of synchrotron radiations which take the form of electromagnetic waves. That is to say; our task is to find the electromagnetic radiations produced by a moving source orbiting near a black hole.

\subsection{Bi-tensors and Green's functions in curved spacetime}

In flat spacetime, the radiations of a moving charged particle have been well-studied\cite{Jackson:1999}. By using the retarded Green's function, the gauge potential caused by a charged particle in motion is simply
\be
A_\mu(x)=4\pi \int d^4 x' G^{\mbox{ret}}_{\mu \alpha}(x-x')J^\alpha(x'),
\ee
where the moving charged particle induces the current
\be
J^\alpha(x')=e\int d\tau Z^\alpha(\tau)\delta^{(4)}(x'-z_0(\tau)).
\ee
$Z^\alpha$ is the 4-velocity of the particle and $z_0(\tau)$ characterize the worldline of the particle. The corresponding gauge potential is the {\it Li\'enard-Wiechert} potential. The corresponding field strength consists of two parts: one part independent of the acceleration is the static field, and the other, depending linearly on the acceleration, is the radiation field.

In curved spacetime, a similar radiation problem has been addressed by DB in \cite{1960Radiation} to understand the radiation damping and the Equivalence Principle for the charged particle in motion. Next, we would like to give a brief and necessary review of the pioneering work by DB. A key point in DB's treatment is to develop the theory of bi-tensors connecting the tensor fields at two different spacetime points, $x$ and $z$, to study the covariant Green's functions. (For a different approach using the vielbein formalism, see \cite{Hobbs:1968}) .

We closely follow DB's conventions. Let $z$ be the point of the source, and $x$ be other points in the spacetime. The letters of the Greek alphabet $\alpha$ to $\kappa$ are assigned to the point $z$, while indices taken from $\lambda$ to $\omega$ are left to the points $x$. For example, $z^\alpha$ and $x^\mu$ are the coordinates of $z$ and $x$, respectively. $s(x,z)$ is defined as the bi-scalar of the geodesic interval, which denotes the geodesic length connecting $x$ and $z$. We can conclude that $\lim \limits_{z\to x}s(x,z)=0$, and we can see that $s(x,z)=ds(x)$ when considering $z=x+dx$, that is to say, at this moment we can take $s$ as the proper time of the geodesic through the point $x$. Moreover, $s>0$ denotes a spacelike interval, and $s<0$ means the interval is timelike, while $s=0$ defines the light cone. Given a sufficiently large interval, $s$ may change sign. However, if we focus on a small neighborhood of a given source, the geodesic interval is single-valued. Thus the covariant expansions can be performed to identify the asymptotic forms of Green's functions containing the information of the electromagnetic waves. It turns out to be convenient to introduce the quantity
\bea\label{delta}
\sigma\equiv\pm1/2s^2\,,
\eea
to avoid the "branch point" problem, which is positive for spacelike intervals and negative for timelike ones.

Furthermore, in addition to the ordinary tensors, which act on the point $z$ or $x$, we need to introduce bi-tensors, of which some indices refer to the point $z$ and the other ones refer to the point $x$. For example, we have a bi-vector $T_{\mu\alpha}$, in which $\mu$ refers to the point $x$ and $\alpha$ refers to the point $z$. Then, it is useful to introduce the bi-vector of geodesic parallel displacement, denoted by $\bar{g}_{\mu\alpha}(x,z)$, which plays the role of transforming the given bi-tensor into a new bi-tensor all of whose indices refer to the same point. The definition of $\bar{g}_{\mu\alpha}$ is given by the following equations
\bea
\nabla^\nu \sigma\nabla_\nu\bar{g}_{\mu\alpha}=0,\quad \nabla^\gamma \sigma\nabla_\gamma\bar{g}_{\mu\alpha}=0,\quad \lim \limits_{x\to z}\bar{g}_\mu^{~\alpha}=\delta_\mu^{~\alpha}. \nn
\eea
For a local vector $V_\mu$ at the point $x$, it can be generated from $V_\alpha$ at the point $z$ by parallel displacement along the geodesic from $z$ to $x$, that is, $ V_\mu=\bar{g}_\mu^{~\alpha}V_\alpha$. Similarly, one can easily apply the geodesic parallel displacement to local tensors of arbitrary order. For our work, we will not show more details of bi-tensors, which can be found in \cite{1960Radiation}.

\begin{figure}[t]
\centering
\includegraphics[width=8cm]{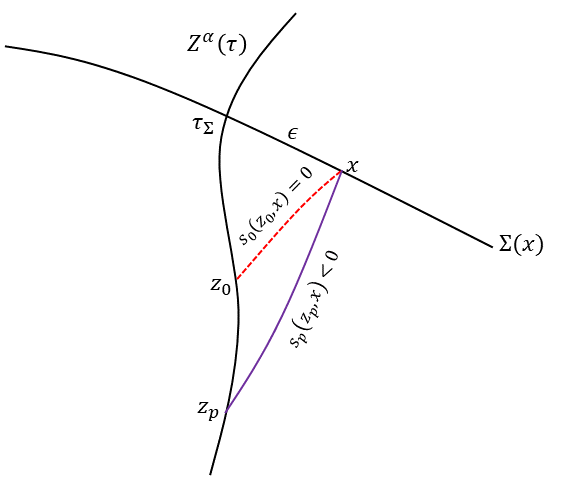}
\caption{A diagram of radiations at the field point $x$ on a spacelike hypersurface $\Sigma(x)$. $Z^\alpha$ is the $4$-velocity of the charged source.}
\label{dia}
\end{figure}

Since we are interested in electromagnetic radiations, we next turn to the solutions of the covariant vector wave equations, which take the form
\bea\label{vwe}
\nabla^\nu\nabla_\nu A_\mu+R_\mu^{~\nu}A_\nu=J_\mu\,,
\eea
in the Lorenz gauge $\nabla_\mu A^\mu=0$. Green's functions for the vector wave functions can be obtained from the Hadamard's "elementary solution" \cite{Hadamard:1923}
\bea
G^{(1)}_{\mu\alpha}=(2\pi)^{-2}\left(u_{\mu\alpha}\sigma^{-1}+v_{\mu\alpha}\log\sigma+w_{\mu\alpha}\right)
\eea
by moving into the complex plane, where, $u_{\mu\alpha}, v_{\mu\alpha}, w_{\mu\alpha}$ are bi-vectors. As a result, the ``symmetric" Green's function $\bar{G}_{\mu\alpha}(x,z)$ can be found in this form\cite{1960Radiation}
\bea\label{uv}
\bar{G}_{\mu\alpha}(x,z)=(8\pi)^{-1}[u_{\mu\alpha}\delta(\sigma)-v_{\mu\alpha}\theta(-\sigma)]
\eea
where $\delta(\sigma)$ is the $\delta$-function and $\theta$ is the Heaviside function with $u_{\mu\alpha}\propto\bar{g}_{\mu\alpha}$ and $v_{\mu\alpha}$ being determined by solving the source free equation $\nabla^\nu\nabla_\nu G^{(1)}_{\mu\alpha}+R_\mu^{~\nu}G^{(1)}_{\nu\alpha}$=0. The term $u_{\mu\alpha}\delta(\sigma)$ shows that a sharp pulse of electromagnetic radiation travels along the null geodesics between the source $z$ and the field $x$, with its polarization vector being parallel-transported. The other term $v_{\mu\alpha}\theta(-\sigma)$ in the Green's function, often referred to as the "tail" term, describes the electromagnetic radiations produced by the source whose geodesic intervals $s(z,x)$ are timelike. It shows that the sharp pulse can develop a "tail" due to the scattering with the "bump" in curved spacetime. Then the retarded Green's function that we desire for the vector wave equation can be formally expressed as
\bea
G^\ret_{\mu\alpha}(x,z)=2\Theta[\Sigma(x),z]\bar{G}_{\mu\alpha}(x,z),
\eea
where $z$ is the source point, $\Sigma(x)$ is an arbitrary spacelike hypersurface containing the field point $x$, and $\Theta[\Sigma(x),z]=1-\Theta[z,\Sigma(x)]=1$ when $z$ lies to the past of $\Sigma(x)$ and $\Theta[\Sigma(x),z]=0$ when $z$ lies to the future. In Fig. \ref{dia}, we show the radiations received at a field point $x$ on $\Sigma(x)$.

\subsection{Electromagnetic fields near a point particle}
The Green's functions $\bar{G}_{\mu\alpha}(x,z)$ cannot be solved precisely in generic curved spacetimes. Nevertheless, we can do covariant expansions of $\bar{G}_{\mu\alpha}(x,z)$ around $z$ to obtain the electromagnetic fields generated from the source in the very near region of point $z$. In particular, since we are interested in the synchrotron radiations emitted from non-geodesic charged particles, we only need to include the contributions from the history of the particle at and before the point $z_0$ to determine the electromagnetic field at point $x$ in Fig. \ref{dia}.
As shown in Fig. \ref{dia}, $Z^\alpha$ is the tangent vector of the world line of the source, and the proper time is set to be $\tau$. For any $\tau$, one always has a certain spacelike hypersurface $\Sigma$ whose normal vector is $Z^\alpha$. Then the induced metric on $\Sigma$ can be defined as $h^{\alpha\beta}=g^{\alpha\beta}+Z^\alpha Z^\beta$. At the proper time $\tau_\Sigma$, the source arrives at the point $z$. We introduce a fixed field point $x$ which has a small spatial distance $\epsilon$ deviated from $z_\Sigma$ on the hypersurface $\Sigma$, and in terms of the induced metric, we can have the following relation:
\bea\label{sigma-epsilon-rela}
h^{\alpha\beta}\nabla_\alpha\sigma\nabla_\beta\sigma=\epsilon^2+\mathcal{O}(\epsilon^3)\,.
\eea
This relation tells us that the quantities like $\nabla_\mu \sigma$ are of order $\mathcal{O}(\epsilon)$. For each geodesic interval $s$ connecting the source point $z$ and the field point $x$, following the results in \cite{1960Radiation}, we can have the expansions of the bi-vectors:
\bea\label{uvf}
u_{\mu\alpha}&=&\left[1-\frac{1}{12}R^{\beta\gamma}\nabla_\beta\sigma\nabla_\gamma\sigma+\mo(\epsilon^3)\right]\bar{g}_{\mu\alpha},\nn\\
v_{\mu\alpha}&=&-\frac{1}{2}\bar{g}^\beta_{~\mu}\left(R_{\alpha\beta}-\frac{1}{6}g_{\alpha\beta}R\right)+\mo(\epsilon)\,,
\eea
and
\bea\label{nablaf}
\nabla_\beta \nabla_\alpha\sigma&=&g_{\alpha\beta}+\frac{1}{3}R_{\alpha~\beta}^{~\gamma~\delta}\nabla_\gamma\sigma\nabla_\delta\sigma+\mathcal{O}(\epsilon^3)\,,\nn\\\nabla_\beta\bar{g}_{\mu\alpha}&=&-\frac{1}{2}R_{\delta\alpha\beta}^{~~~\gamma}\nabla_\gamma\sigma+\mo(\epsilon^2)\,,\nn\\
\nabla_\beta u_{\mu\alpha}&=&\left(-\frac{1}{2}\bar{g}^{\delta}_{~\mu}R_{\delta\alpha\beta}^{~~~\gamma}-\frac{1}{6}\bar{g}_{\mu\alpha}R^{\gamma}_{~\beta}\right)\nabla_\gamma\sigma+\mo(\epsilon^2). \nn
\eea
Here we emphasize that the so-called higher order terms $\mo(\epsilon^2)$ and $\mo(\epsilon^3)$ mean the related terms involve two and three covariant derivatives of $ \sigma $, respectively. Thus, the above expansions are valid for any geodesic interval $s$, even for the geodesic intervals on the light cone with $s=0$.

With the expansions of Green's functions, the retarded gauge potential of a point charged particle reads
\bea
A^\ret_\mu&=&4\pi e \int_{-\infty}^{+\infty}G^\ret_{\mu\alpha}Z^\alpha d\tau\nn\\
&=&-e\left[u_{\mu\alpha}Z^\alpha(\nabla_\beta\textcolor{red}{\sigma} Z^\beta)^{-1}\right]_{\tau=\tau_0}+e\int_{-\infty}^{\tau_0}v_{\mu\alpha}Z^\alpha d\tau\nn
\eea
where we have defined $\tau_0$ as the proper time of the point particle at the point $z_0$ in Fig. \ref{dia}, and $e$ is the particle's charge. The corresponding
electromagnetic field strength tensor can be obtained by straightforward differentiation
\bea\label{eof}
F^\ret_{\mu\nu}&=&2e[Z^\alpha \nabla_{[\mu}\sigma u_{\nu]\alpha}(Z^\beta Z^\gamma\nabla_\beta\nabla_\gamma \sigma )(D_\tau \sigma)^{-3}\nn\\
&&-(Z^\alpha D_\tau \nabla_{[\mu}\sigma u_{\nu]\alpha}+D_\tau Z^\alpha \nabla_{[\mu}\sigma u_{\nu]\alpha})(D_\tau \sigma)^{-2}\nn\\
&&+Z^\alpha(\nabla_{[\mu} u_{\nu]\alpha}+\nabla_{[\mu}\sigma v_{\nu]\alpha})(D_\tau\sigma)^{-1}]_{\tau=\tau_0}\nn\\
&&-2e\int_{-\infty}^{\tau_0}d\tau Z^\alpha\nabla_{[\mu} v_{\nu]\alpha}
\eea
where we have introduced the derivative operator along the vector $Z^\alpha$, that is, $D_\tau\equiv Z^\beta\nabla_\beta$. In Eq. (\ref{eof}), the leading terms of $F^\ret_{\mu\nu}$ are of order $\mo(\epsilon^{-2})$, and they are, in fact, from the Coulomb potential, which would give a renormalization of the mass of the point particle but won't contribute to the radiations. The subleading terms being of order $\mo(\epsilon^{-1})$ from the lightlike part would contribute to the radiations mostly. The synchrotron radiations originating from the gravity involving curvature tensors are of order $\mo(\epsilon^0)$ both in the lightlike part and in the "tail" part. The radiations from the complicated "tail" term involve the integrations over the whole history of the particle, and they won't be observed if we only keep them for a limited time interval. Moreover, such radiations are sub-leading but could be significant in a strong gravitational field.

\subsection{Synchrotron radiations and polarization vector}\label{sec:srpv}

Next, we turn to find out the electromagnetic radiations and synchrotron radiations generated by a non-geodesic point particle acted upon by the Lorentz force and extract the corresponding electric component as the polarization vector, as well as its strength.

Note that the geodesic interval between the point $z_0$ and the point $x$ is on the light cone; thus, we are allowed to define $\nabla_\alpha\sigma\equiv \epsilon k_\alpha$ at the point $z_0$, where $k_\alpha k^\alpha=0$ is a null vector along the light cone. Then, it's easy to conclude that at the point $x$, we have $\nabla_\mu\sigma=-\bar{g}_\mu^{~\alpha}\nabla_\alpha \sigma=-\epsilon k_\mu$. After some calculations, the electromagnetic field tensor in Eq. (\ref{eof}) can be expanded and rewritten as
\bea\label{elege}
F^\ret_{\mu\nu}=\sum_{d=-2}^{\infty} F^{(d)}_{\mu\nu} \epsilon^d+(\text{tail part})\,,
\eea
where
\bea
F^{(-2)}_{\mu\nu}&=&2e\left[Z^\alpha k_{[\mu}\bar{g}_{\nu]\alpha}(k_\eta Z^\eta)^{-3}\right]_{\tau=\tau_0}\,\\
F^{(-1)}_{\mu\nu}&=&4e\left[k_{[\nu}\bar{g}_{\mu]\alpha}( k_\beta D_\tau Z^{[\beta} Z^{\alpha]})(k_\eta Z^\eta)^{-3}\right]_{\tau=\tau_0}\,\label{fmn1}\\
F^{(0)}_{\mu\nu}&=&2e\bigg\{k_{[\nu}\bar{g}_{\mu]\alpha}Z^\alpha\left(\frac{1}{3}R_{\gamma~\beta}^{~\theta~\delta}k_\theta k_\delta Z^\beta Z^\gamma+\frac{1}{12}R^{\beta\gamma}k_\beta k_\gamma\right)(k_\eta Z^\eta)^{-3}\nn\\
&&+\left(\frac{1}{2}k_{[\nu}\bar{g}_{\mu]}^{~\delta}R_{\delta\alpha\beta}^{~~~\gamma}+\frac{1}{6}k_{[\nu}\bar{g}_{\mu]\alpha}R_\beta^{~\gamma}\right)k_\gamma Z^\alpha Z^\beta(k_\eta Z^\eta)^{-2}\nn\\
&&+\frac{1}{6}\bar{g}^\theta_{~[\nu}\bar{g}_{\mu]\alpha}R_{\theta~\beta}^{~\delta~\gamma}k_\gamma k_\delta Z^\alpha Z^\beta(k_\eta Z^\eta)^{-2}\nn\\
&&-Z^\alpha\bigg[k_\gamma\left(\frac{1}{2}\bar{g}_{~[\nu}^{\beta}\bar{g}_{\mu]}^{~\delta}R_{\delta\alpha\beta}^{~~~\gamma}+\frac{1}{6}\bar{g}_{~[\nu}^{\beta} \bar{g}_{\mu ] \alpha} R_\beta^{~\gamma}\right)\nn\\
&&+\frac{1}{2} k_{[\nu}\bar{g}_{\mu]}^{~\beta}\left(R_{\alpha\beta}-\frac{1}{6}g_{\alpha\beta}R\right)\bigg](k_\eta Z^\eta)^{-1}\bigg\}_{\tau=\tau_0}
\eea
In Eq. (\ref{elege}), the electromagnetic radiations of charged particles involve a tail term which gets contribution only from the spacetime curvature. Due to the presence of a tail term, the EEP is invalid for the electromagnetic radiations of charged particles in curved spacetime. Nevertheless, as the leading order of the tail part is of $\mo(\epsilon^0)$, if we focus on the photons with high enough frequency, the tail part can be dropped and only $F^{(-2)}_{\mu\nu}$ and $F^{(-1)}_{\mu\nu}$ terms survive. As a result, the EEP is valid for electromagnetic radiations of charged particles under the approximation of geometrical optics. Our results are consistent with Eq. (23) in QW's paper \cite{Quinn:1996am}.

Moreover, the term $F^{(-2)}_{\mu\nu}$ corresponds to the electromagnetic field generated by the Coulomb potential. And a nonvanishing $F^{(-1)}_{\mu\nu}$ would give a non-trivial electromagnetic radiations dependent on the $4$-acceleration of charged particles $D_\tau Z^\alpha$. In other words, if the charged particles move along geodesic trajectories, this term would vanish, so we focus on the particles in non-geodesic motions in the following discussion.

When charged particles of mass $m$ are only impacted by a magnetic field
\bea\label{Lorentzforce}
D_\tau Z^\alpha=\frac{e}{m}F^\alpha_{~\beta}Z^\beta\,,
\eea
electromagnetic radiations are the so-called synchrotron radiations, which have an excellent property of polarization encoding the information of the source and the magnetic field. Therefore, it is fascinating to have a careful study of synchrotron radiations.

The electric vector can be extracted from the electric components in $F^{(-1)}_{\mu\nu}$ via a timelike tetrad
\bea \label{ftopol}
E^\mu\equiv\frac{1}{\epsilon} F^{(-1)\mu\nu}\xi_{\nu}\,,
\eea
where $\xi_\nu$ is the timelike component of a local Minkowski frame that can be chosen arbitrarily.
The form of $F^{(-1)}_{\mu\nu}$ of a general electromagnetic wave can always be rewritten as
\bea \label{sstrr}
F^{(-1)}_{\mu\nu}\equiv k_\mu \mathcal{A}_{\nu}-k_\nu \mathcal{A}_{\mu}\,,
\eea
where $\mathcal{A}_{\mu}$ can be read from the expression (\ref{fmn1})
\bea\label{aex}
\mathcal{A}_{\mu}=-4e\left[\bar{g}_{\mu\alpha}( k_\beta D_\tau Z^{[\beta} Z^{\alpha]})(k_\eta Z^\eta)^{-3}\right]_{\tau=\tau_0}.
\eea
Taking into account the conditions, $k\cdot k=0$ and $k\cdot \mathcal{A}=0$, we obtain the luminosity of the synchrotron radiations emitted by the source $Z^\alpha$:
\bea
L=4\pi\epsilon^2 E^\mu E_\mu=4\pi(k_\rho Z^\rho)^2 (\mathcal{A}_\sigma \mathcal{A}^\sigma)\,,
\eea
Plunging the definition of $k_\alpha$ into the Eq. (\ref{sigma-epsilon-rela}) we have $k_\alpha Z^\alpha=1$. Combining with this relation, we substitute the expression of $\mathcal{A}_\mu$ into $L$ and finally find the
luminosity
\bea\label{inten}
L=4\pi|k_\beta D_\tau Z^\beta Z^\alpha- D_\tau Z^\alpha|^2_{\tau=\tau_0}.
\eea

Next, we move to the polarization vector of the synchrotron radiations. Because the polarization vector $f^\mu\propto E^\mu$ should be normalized and transverse along the null geodesic, we have $f^\mu f_\mu=1\,, f^\mu k_\mu=0$. Note that a general polarization vector of the electromagnetic wave has four components but only one physical degree of freedom. The reason is that apart from the above two constraints, gauge freedom exists. We can see that the physical information of the polarization vector remains unchanged if one chooses different observers. Considering the fact that $\xi^\mu$ can be regarded as the $4$-velocity of an observer, we change $\xi$ to $\xi^\prime$, it is not hard to check that the variation $\delta f^\mu\equiv f^{\prime\mu}-f^\mu$ satisfies $
\delta f^\mu \delta f_\mu=0\,, \delta f^\mu k_\mu=0$, which means that it is proportional to the $4$-momentum of radiation, i.e., $\delta f^\mu \propto k^\mu$. In this sense, $f^{\prime\mu}$ is indistinguishable from $f^\mu$: they differ only by a gauge transformation proportional to $k^\mu$. Moreover, from Eq. (\ref{sstrr}), we can see that
\bea\label{fmu}
f^\mu \sim F^{(-1)\mu\nu}\xi_{\nu}=(\mathcal{A}_\nu\xi^\nu)k_\mu-(k_\nu\xi^\nu)\mathcal{A}_\mu\,.
\eea
The first term $(\mathcal{A}_\nu\xi^\nu)k_\mu\propto k_\mu$ is just a gauge term that can be dropped, and therefore without loss of generality, the polarization vector can be taken as
\bea\label{ffeq}
f_\mu=\frac{\mathcal{A}_\mu}{\sqrt{\mathcal{A}^2}}=N^{-1}\left[\bar{g}_{\mu\alpha}( k_\beta D_\tau Z^{[\beta} Z^{\alpha]})\right]_{\tau=\tau_0}\,,
\eea
where we have normalized the vector appropriately by a factor $N^{-1}$. Here we would like to stress that one can recover the term $(\mathcal{A}_\nu\xi^\nu)k_\mu\propto k_\mu$ in Eq. (\ref{fmu}) from the expression in Eq. (\ref{ffeq}) under a certain observer denoted by $\xi_\mu$ by requiring
\bea
f^\mu \xi_\mu=0\,,
\eea
which can be easily derived by the definition of polarization vector $f^\mu$.

The formulas \eqref{inten} and \eqref{ffeq} are the main results of our work. They may reduce to those in flat spacetime, according to the EEP. More importantly, they can be applied to explore the polarized structure of synchrotron radiations in curved spacetime without resorting coordinate transformations.

\section{Polarization of synchrotron radiations around a Schwarzschild black hole}\label{sec3}

As an application of Eq. (\ref{ffeq}), let us study the polarized image of synchrotron radiations in the Schwarzschild black hole spacetime. The Schwarzschild metric takes the form
\bea\label{mmetricc}
ds^2=-\left(1-\frac{2M}{r}\right)d t^2+\left(1-\frac{2M}{r}\right)^{-1}d r^2+r^2 (d\theta^2+\sin^2\theta d\phi^2)\,,
\eea
where $M$ is the mass of the black hole. For simplicity, we assume the black hole is immersed in a uniform magnetic field \cite{Wald:1974np} with a nonzero component of the gauge field \footnote{The analysis of the polarized images employed by our method has been applied to a Melvin magnetic field in \cite{Zhu:2022amy}.}
\bea
A_\phi=\frac{1}{2}B r^2 \sin^2\theta\,,
\eea
where $B$ is a constant standing for the strength of the uniform magnetic field. In particular, we want to emphasize that the model we consider here does not involve the backreaction of the magnetic field to the background spacetime. The spacetime is still characterized by the Schwarzschild black hole metric.

\subsection{The motion of source}
\begin{figure}[t]
\centering
\includegraphics[width=8cm]{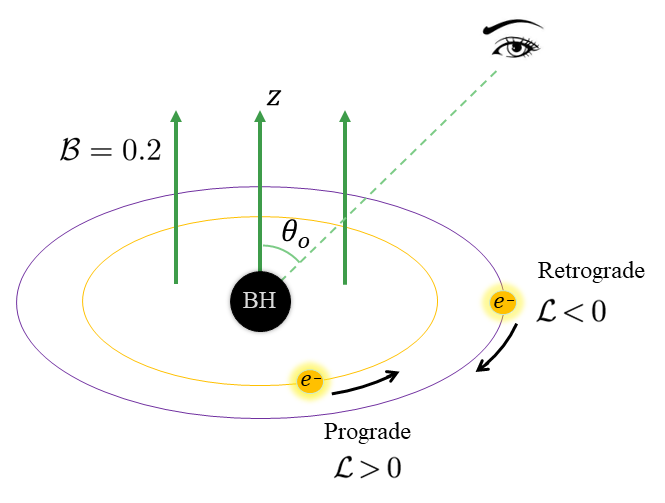}
\caption{A diagram of retrograde ($\mathcal{L}<0$) and prograde ($\mathcal{L}>0$) orbits of electrons. We set the direction of the magnetic field $\vec{\mathcal{B}}$ along the positive axis of $z$. $\theta_o$ is the inclination angle of the observer.}
\label{dia2}
\end{figure}

In this work, the source of our interest is composed of the large number of electrons moving in the region near a Schwarzschild black hole. Since the Schwarzschild black hole is bathed in a uniform magnetic field, the equation of motion of electrons must include a Lorentz force term\cite{Frolov:2010mi}. For such an equation, one cannot obtain a general analytical solution. To simplify our discussions, we will constrain the charged particles' motion to the equatorial plane vertical to the uniform magnetic field and consider their circular orbits. Thus, the tangent vector of the world line of the source takes
\bea
Z^\alpha=(\dot{t}_s, \dot{r}_s, 0, \dot{\phi}_s)\,,
\eea
where we have set $\theta_s=\pi/2, r_s=\text{const}$ with the subscript $s$ signifying the source, and the $\cdot$ means the derivative with respect to the proper time $\tau$. Because the energy $E=-mZ_t$ and angular momentum $L=mZ_\phi+qA_\phi$ are conserved along the world line of the source, where $m$ and $q$ are the mass and the charge of the source, respectively, we can obtain
\bea\label{rsfn}
\dot{r}_s^2=\mathcal{R}(r_s)\equiv\mathcal{E}^2-\left(1-\frac{2M}{r_s}\right)+\left[1+r_s^2\left(\frac{\mathcal{L}}{r_s^2}-\mathcal{B}\right)^2\right]=0\,,
\eea
from the normalization condition of $4$-velocity, that is, $Z^\alpha Z_\alpha=-1$. Note that in Eq. (\ref{rsfn}), we reparameterize $B, E, L$ as
\bea
\mathcal{B}&\equiv&\frac{q B}{2 m} \,,\nn\\
\mathcal{E}&\equiv&E/m=\dot{t}_s\left(1-\frac{2M}{r_s}\right) \,,\nn\\
\mathcal{L}&\equiv&L/m=\dot{\phi}_s r_s^2 + \mathcal{B}r_s^2 \,.
\eea
In our convention, $\mathcal{B}$ is always set to be positive, as a negative $\mathcal{B}$ system is equivalent to a positive $\mathcal{B}$ system via a reflection transformation.
Thus, a positive $\mathcal{L}$ corresponds to a prograde orbit and a negative one to a retrograde orbit, see Fig. \ref{dia2}. For circular orbits, one further has $\mathcal{R}^\prime(r_s)=0$ corresponding to $\ddot{r}_s=0$, with `` $\prime$ '' the derivative with respect to $r$, which gives
\bea\label{eeemm}
r_s\left(\frac{\mathcal{L}}{r_s^2}-\mathcal{B}\right)\left[\frac{\mathcal{L}}{r_s^2}\left(1-\frac{3M}{r_s}\right)+\mathcal{B}\left(1-\frac{M}{r_s}\right)\right]-\frac{M}{r_s^2}=0\,.
\eea
Note that $\mathcal{E}$ does not appear in Eq. (\ref{eeemm}). Thus we can use this equation to identify the reparameterized angular momentum $\mathcal{L}$ for given $r_s$ and $\mathcal{B}$. Then, plunging $\mathcal{L}$, $\mathcal{B}$ and $r_s$ in Eq. (\ref{rsfn}), one can obtain the reparameterized energy $\mathcal{E}$. Then the $4$-vector $Z^\alpha$ can be completely determined. The initial conditions we need to know are the source orbit's radius and the magnetic field's strength around the black hole.

In addition, only some of the timelike circular orbits are stable. For example, for a vacuum Schwarzschild black hole, the timelike circular orbits inside $r=6M$ are unstable, and the critical radius is known as the innermost stable circular orbit (ISCO). In the following, we would like to focus on the situation in the radial position of the source located from $r_s=6M$ to $r_s=10M$.

\subsection{Propagation of light and image of source}

Up to now, we have known the information on the orbits of the source and the polarization vector of synchrotron radiations caused by the source. However, since our final task is to find out the information on the polarization of synchrotron radiations at the position of an observer, which is generally placed at infinity, we have to figure out the propagation of electromagnetic waves from the source to the observer. For simplicity, we only consider high-frequency emissions such that we can use the geometric optics approximation to deal with the propagation problem. That is to say, the emissions that spread away from the source go along null geodesics in spacetime.

Null geodesics in Schwarzschild black hole spacetime have been studied very clearly\cite{Gralla:2019drh}. Here we only introduce some basic physical quantities and useful formulas. In our convention, we use $k^\mu$ to denote the $4$-momentum of photons. Even though there is spherical symmetry in the spacetime of a Schwarzschild black hole such that the source can be fixed at $\theta_s=\pi/2$, the observers are not necessarily on the equatorial plane. Thus $k^\theta$ cannot be set to zero in general. Along the null geodesics, there are three conserved quantities, the energy $\omega$ and the angular-momentum $l$ and Carter constant $Q^2$,
\bea
\omega=-k_t\,,\quad l=k_\phi\,,\quad Q^2=k_\theta^2+\frac{k_\phi^2}{\sin^2\theta}.
\eea
Then, from $k^\mu k_\mu=0$, one can find
\bea\label{neom}
k^t/\omega=\left(1-\frac{2M}{r}\right)^{-1}\,,\, k^r/\omega=\pm\frac{\sqrt{R(r)}}{r^2}\,,\, k^\theta/\omega=\pm\frac{\sqrt{\Theta(\theta)}}{r^2}\,,\, k^\phi/\omega=\frac{\lambda}{r^2\sin^2\theta}\,,
\eea
where,
\bea
R(r)&=&r^4-(r^2-2Mr)\rho^2\,,\\
\Theta(\theta)&=&\rho^2-\frac{\lambda^2}{\sin^2\theta}\,
\eea
with the impact parameters $\lambda\equiv l/\omega$ and $\rho^2\equiv Q^2/\omega^2$.
Next, consider an observer with the coordinate $(t_o, r_o, \theta_o, \phi_o)$. The position of the photons reaching the eyes of the observer can be described in terms of celestial coordinates, which are usually chosen as
\bea\label{impaa}
\alpha&=&-\frac{\lambda}{\sin\theta_o}\,,\nn\\
\beta&=&\pm_o\sqrt{\Theta(\theta_o)}\,,
\eea
where $\pm_o=\text{sign}(k^\theta_o)$ denotes the sign of $k^\theta$ in the observer's frame. Alternatively, for convenience, one can use the polar coordinates $(\rho, \varphi)$ instead of the Cartesian coordinates $(\alpha, \beta)$. The two coordinate frames are related by
\bea\label{varphid}
\rho=\sqrt{\alpha^2+\beta^2}=\frac{Q}{\omega}\,,\quad\quad \tan\varphi=\frac{\beta}{\alpha}\,.
\eea
Then, considering the null geodesics connecting the source $(t_s, r_s, \pi/2, \phi_s)$ to the observer $(t_o, r_o, \theta_o, \phi_o)$, one integrate Eq. (\ref{neom}) and has
\bea
I_r=G_\theta\,,\quad\Delta\phi=\phi_o-\phi_s=\lambda G_\phi\,,
\eea
where
\bea
I_r=\fint^{r_o}_{r_s}\frac{dr}{\pm_r\sqrt{R(r)}}\,,\quad G_\theta=\fint_{\pi/2}^{\theta_o}\frac{d\theta}{\pm_\theta\sqrt{\Theta(\theta)}}\,,\quad G_\phi=\fint_{\pi/2}^{\theta_o}\frac{\csc^2\theta d\theta}{\pm_\theta\sqrt{\Theta(\theta)}}
\eea
with the notation $\fint$ indicating that these integrals are path integrals along the photon trajectories. And the symbols $\pm_r=\text{sign}(k^r)$ and $\pm_\theta=\text{sign}(k^\theta)$ indicate the sign of $k^r$ and $k^\theta$, which switch at radial and angular turning points, respectively. Let $\bar{m}$ be the number of times light rays cross the equatorial plane from the source to the observer. One can find
\bea\label{gtheta}
G_\theta=\frac{1}{\rho}\arccos\left(-\frac{\sin\varphi}{\sqrt{\sin^2\varphi+\cot^2\theta_o}}\right)+\frac{\bar{m}\pi}{\rho}\,.
\eea
Here and hereafter, we set $M=1$ for simplicity and without loss of generality. In addition, different values of $\bar{m}$ correspond to the $(\bar{m}+1)\text{th}$ image on the observers' screen. In this work, we would like to focus on the primary image and secondary image of the source. Now, from Eqs. (\ref{impaa}-\ref{gtheta}), one can completely obtain the image of the source on the observer's screen, given the spacetime coordinates of the source and the observer.

\subsection{Penrose-Walker constant}
In this subsection, we study polarization information transmission along null geodesics connecting the source and the observer. Note that, for a spacetime of Petrov type-D \cite{Lupsasca:2018tpp}, there is a conserved quantity along the geodesics for a parallel-transported and transverse vector, dubbed as the Penrose-Walker(PW) constant, which is typically written as
\bea\label{pwc}
\kappa=2r[(k^\mu \hat{l}_\mu)(f^\nu \hat{n}_\nu)-(k^\mu \hat{m}_\mu)(f^\nu\hat{\bar{m}}_\nu)]\equiv \omega\left(\kappa_1+i \kappa_2\right)\,,
\eea
for a Schwarzschild black hole spacetime. In the above formula, $r$ is the radial coordinate introduced in the metric (\ref{mmetricc}), $k^\mu$ is the $4$-momentum of photons, $f^\mu$ is the polarization vector, $\kappa_{1,2}$ are rescaled real part and imaginary part of PW constant, and $\{\hat{l}^\mu, \hat{n}^\mu, \hat{m}^\mu, \hat{\bar{m}}^\mu\}$ are the Newman-Penrose tetrads taking the form
\bea
\hat{l}&=&\frac{1}{\sqrt{2(1-\frac{2}{r})}} \left[\partial_t+(1-\frac{2}{r})\partial_r \right] \,,\nn\\ \hat{n}&=&\frac{1}{\sqrt{2(1-\frac{2}{r})}} \left[\partial_t-(1-\frac{2}{r})\partial_r \right] \,,\nn\\
\hat{m}&=&\frac{1}{\sqrt{2} r} \left[\partial_{\theta}+\frac{i}{\sin\theta}\partial_{\phi} \right]\,,\nn\\
\hat{\bar{m}}&=&\frac{1}{\sqrt{2} r} \left[\partial_{\theta}-\frac{i}{\sin\theta}\partial_{\phi} \right]\,.
\eea
From Eq. (\ref{pwc}), we can see that all the information of polarization is encoded in the PW constant at the source, thus considering the light rays connecting the source and the observer, we can decode the information of polarization at the observer. On the screen of observers, the polarization vector can be read from the PW constant as
\bea\label{pollf}
\vec{\mathcal{E}}&=&(\mathcal{E}_\alpha,\mathcal{E}_\beta)\,,\nn\\
&=&\frac{1}{\rho^2} (\beta \kappa_2+\alpha \kappa_1,\beta\kappa_1-\alpha\kappa_2).
\eea

\subsection{Image fluxes}

In this subsection, we turn our attention to the calculations of the image fluxes. Since the source in this work is a charged particle with a finite size, Eq. (\ref{inten}) cannot be directly used to give the total flux of the image on the screen. The formalism for dealing with such a problem was developed in \cite{Cunningham:1973b} for an extremal black hole and was generalized to include the non-extreme case in \cite{Gralla:2017ufe}. In this paper, we closely follow the method presented in \cite{Gralla:2017ufe}, which has been applied to calculate the observational signature for near-extremal black holes in the modified theory in \cite{Guo:2018kis, Guo:2019lur}.

Firstly, we introduce the Newtonian flux, which is related to the luminosity by
\bea
F_N=\frac{L}{4\pi r_o^2},
\eea
where $r_o$ is the distance between the observer and the source. The distance $r_o$ can be large, $r_o\gg 1$. Then, we assume that the source has a minimal proper radius $b\ll 1$ emitting with the intensity $I_s$ in the rest frame of the source. The relation between the intensity and the luminosity is given by
\bea
I_s=\frac{L}{4\pi^2 b^2}.
\eea
Taking the gravitational effect into account, we find the fluxes of the image
\bea
F_o=\oiint \frac{d\alpha d\beta}{r_o^2}g^4 I_s=\frac{\mathcal{A}}{\pi b^2}g^4 F_N,
\eea
where $g=\nu_o/\nu_s$ is the redshift factor and $\mathcal{A}=\oiint d\alpha d\beta$ is the area of the image.

To determine the area of the image $\mathcal{A}$, we would like to introduce a local Minkowski coordinate system $(T, X, Y, Z)$ in the neighborhood of the source so that we have
\bea
T e_{(t)}+X e_{(r)}+Y e_{(\phi)}-Z e_{(\theta)}=(x^\mu-x_\star^\mu)\partial_\mu,
\eea
where $x^\mu_\star$ are the position coordinates of the source and $e_{(i)}$ are the tetrad of the source. In this work we choose $e_{(i)}$ to be
\bea
e_{(t)}=\gamma\frac{\partial_t+\Omega_s \partial_\phi}{\sqrt{-g_{t t}}},\quad
e_{(r)}=\frac{\partial_r}{\sqrt{g_{r r}}},\quad
e_{(\theta)}=\frac{\partial_\theta}{\sqrt{g_{\theta \theta}}},\quad
e_{(\phi)}=\gamma\left(v_s\frac{\partial_t}{\sqrt{-g_{t t}}}+\frac{\partial_\phi}{\sqrt{g_{\phi \phi}}}\right),
\eea
with $\Omega_s=\frac{d \phi}{d t}$ being the angular velocity of the source and
\bea
\gamma=\frac{1}{\sqrt{1-v_s^2}},\qquad v_s=\sqrt{\frac{g_{\phi \phi}}{-g_{t t}}}\Omega_s.
\eea

Using the terminology introduced in \cite{Gralla:2017ufe}, we call the surface $T=X=0$ the "source screen" and denote by $(Y_s, Z_s)$ the position of the intersection of a light ray with the source screen. Then the area of the image becomes
\bea
\mathcal{A}=\oiint d\alpha d\beta=\left|\frac{\partial(\alpha,\beta)}{\partial(Y_s,Z_s)}\right|\oiint d Y_s d Z_s\,,
\eea
where $\left|\frac{\partial(\alpha,\beta)}{\partial(Y_s,Z_s)}\right|$ is the Jacobian determinant between $(\alpha, \beta)$ and $(Y_s, Z_s)$. In addition, the projection of the hemisphere of the source onto the screen is an ellipse with an area
\bea
\oiint d Y_s d Z_s=\frac{\pi b^2}{|\hat{k}\cdot\hat{X}|},
\eea
where the unit vector $\hat{k}$ is given by
\bea
\hat{k}=\frac{1}{k^{(t)}}\left(k^{(r)}\hat{X}+k^{(\phi)}\hat{Y}-k^{(\theta)}\hat{Z}\right).
\eea
Then we can read the area of the image
\bea
\mathcal{A}=\frac{\pi b^2}{|\hat{k}\cdot\hat{X}|}\left|\frac{\partial(\alpha,\beta)}{\partial(Y_s,Z_s)}\right|\,,
\eea
and find that the image flux takes the form
\bea
F_o=\frac{g^4L}{4\pi r_o^2|\hat{k}\cdot\hat{X}|}\left|\frac{\partial(\alpha,\beta)}{\partial(Y_s,Z_s)}\right|\,.
\eea
Here we omit the detail of the computation of the Jacobian determinant $\left|\frac{\partial(\alpha,\beta)}{\partial(Y_s, Z_s)}\right|$, and suggest the interested readers find the details in \cite{Gralla:2017ufe}.

Hereto, we have completed all the theoretical preparations and are ready to apply our model to some specific situations.

\section{Specific cases}\label{sec4}
In this section, based on the model we introduced, we would like to show some specific cases, including the sources in different orbits, the background with different magnetic field strengths, etc. From Eqs. (\ref{Lorentzforce}) and (\ref{ffeq}), we can see that the strength of the magnetic field $B$ does not influence the direction of the polarization vector. Thus we fix $B=0.2$ in the following discussion.

\begin{figure}[h!]
\centering
\includegraphics[width=15cm]{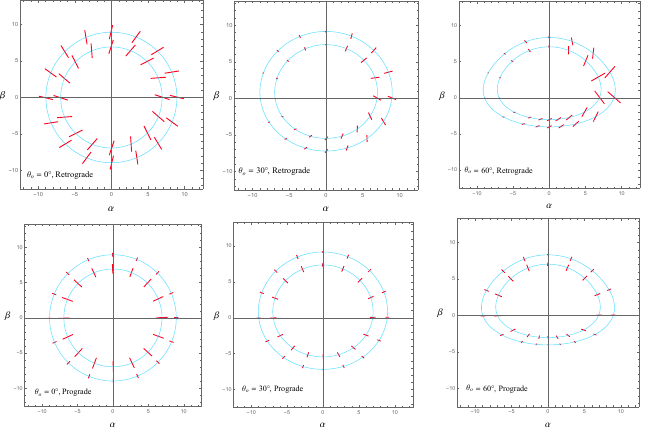}
\caption{Primary images of sources and their polarizations at $\mathcal{B}=0.2$.}
\label{diskp}
\end{figure}

In Fig. \ref{diskp}, we show the primary images of different sources and their polarization directions for $\theta_o=0, \pi/6, \pi/3$ at $\mathcal{B}=0.2$. The upper panel is for the retrograde orbits, and the lower one is for the prograde orbits. The light blue curves are the images of the source, and the red lines stand for the polarizations at different positions. As well, the radii of the source orbits are $r_s=6, 8$ from the inside out. Note that the magnitudes of the image fluxes for different observational angles are very different, so we normalize the image fluxes in specific and suitable ways for each observational angle to show better the information of the fluxes in the plots. More precisely, for $\theta_o=0$, we set the length of the red line $\Len$ to be in direct proportion to the magnitudes of the image fluxes, that is, $\Len\propto F_o$. For $\theta_o=\pi/6$, we set $\Len=\frac{1}{2}\log\frac{F_o}{F_o^\mi}$, where $F_o^\mi$ means the minimum of $F_o$ of all the calculated polarizations for a fixed observational angle $\theta_o$ while for $\theta_o=\pi/3$, we let $\Len=\log \frac{F_o}{F_o^\mi}$.

\begin{figure}[h!]
\centering
\includegraphics[width=15cm]{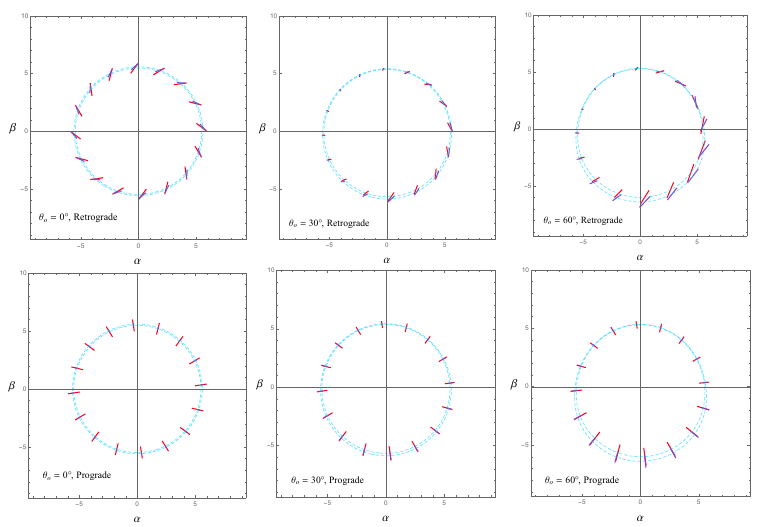}
\caption{Secondary images of sources and their polarizations at $\mathcal{B}=0.2$.}
\label{disks}
\end{figure}

We also show the results of the secondary images of the sources and polarization directions for $\theta_o=0, \pi/6, \pi/3$ at $\mathcal{B}=0.2$ in Fig. \ref{disks}. Likewise, the upper panel shows the images of retrograde orbits, and the lower one shows the pictures of the prograde orbits with light blue curves indicating the photos of the trajectories at $r_s=6, 8$. Different from the primary images, we use the red and blue line segments to present the polarizations for $r_s=6$ and $r_s=8$, respectively, since the pictures of these two orbits are very close and parts of the line segments of the polarizations overlap so that they are hard to distinguish with the same colors. In addition, the normalization of the fluxes in the secondary images is the same as those in the primary ones.

From Fig. \ref{diskp} and Fig. \ref{disks}, we can see qualitatively that for the prograde orbits, the angles between polarization directions and the corresponding lines through the origin and the celestial coordinates are tiny. In other words, the polarizations from the prograde orbits are more likely pointing to the center of the screen of observers. While for the retrograde orbits, the angles between polarization directions and the corresponding lines through the origin and the celestial coordinates become larger as the inclination of the observer $\theta_o$ increases in $[0, \pi/2]$. As for the magnitudes of the image fluxes, there are many differences between the retrograde and prograde orbits for both the primary and secondary images. For the direct images, the magnitudes of the image fluxes change little when $r_s$ goes from $6$ to $8$ for the retrograde orbits, while they decrease significantly for the prograde orbits. For the secondary images, the situation becomes more complicated. In short, the magnitudes of the image fluxes have changed for both the retrograde and prograde orbits, but the changes are more significant for the prograde ones. Moreover, we can find that the Doppler effect is more pronounced for the retrograde orbits when $\theta_o=\pi/6, \pi/3$.

In particular, we can see three patterns of the polarizations in Fig. \ref{diskp} and Fig. \ref{disks}, that is the closed whorl, the open whorl, and the radial pattern. And the patterns of polarizations are quite different for the prograde and retrograde orbits, and are also different for the primary and secondary images. In Fig. \ref{diskp}, for the primary images of the sources in retrograde orbits, the rings of polarization ticks look like closed whorls  when $\theta_o=0^\circ$, but then look like open whorls with the breaking points at $(\alpha=\alpha_{\ma}, \beta=0)$ when $\theta_o=30^\circ$ and $\theta_o=60^\circ$. In Fig. \ref{disks}, for the secondary images of sources in retrograde orbits, the rings of polarization ticks are all closed whorls when $\theta_o=0^\circ, 30^\circ, 60^\circ$. As for the prograde sources, we can see that all the rings of polarization ticks are radial patterns for both the primary and secondary images, no matter what the observation angle is.

\begin{figure}[t]
\centering
\includegraphics[width=8cm]{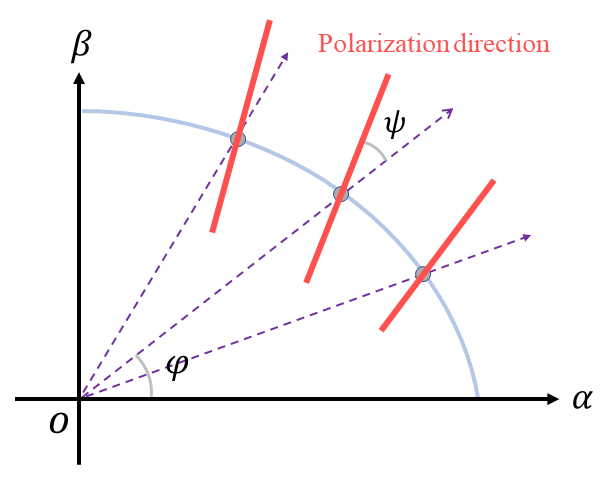}
\caption{A diagram of an angle between the polarization vector and the radial vector at $(\rho, \varphi)$ in the $\alpha-\beta$ plane.}
\label{dia3}
\end{figure}

To make the description of the polarization angle more intuitive, from Eqs. (\ref{varphid}) and (\ref{pollf}) we introduce a new angle defined as
\bea
\psi\equiv-\arctan\left(\frac{\kappa_2}{\kappa_1}\right)+\frac{\text{sign}(\kappa_1)-1}{2}\pi\,.
\eea
Here we specify that the angle $\psi$ increases along the counterclockwise direction; for example, in Fig. \ref{dia3}, we have $\psi>0$. Note that the term $\frac{\text{sign}(\kappa_1)-1}{2}\pi$ is added to ensure the $\psi$ function is continuous in $(-\pi,\pi)$, since $\arctan$-function is usually defined in $[-\pi/2, \pi/2]$. Also, this term doesn't affect the polarization information since $\psi$, and $\psi-\pi$ indicate the same polarized light in physics. Thus, we can see that after the vector at the angle of $\varphi\in[0, 2\pi]$ from the $\alpha$-axis goes through a $\psi$ rotation, and it's going to be parallel to the polarization vector at the point $(\rho, \varphi)$ on the $\alpha-\beta$ plane. Now, we are allowed to use $\psi$ to quantitatively reflect the extent to which the polarization direction deviates from $\varphi$.

\begin{figure}[h!]
\centering
\includegraphics[width=15cm]{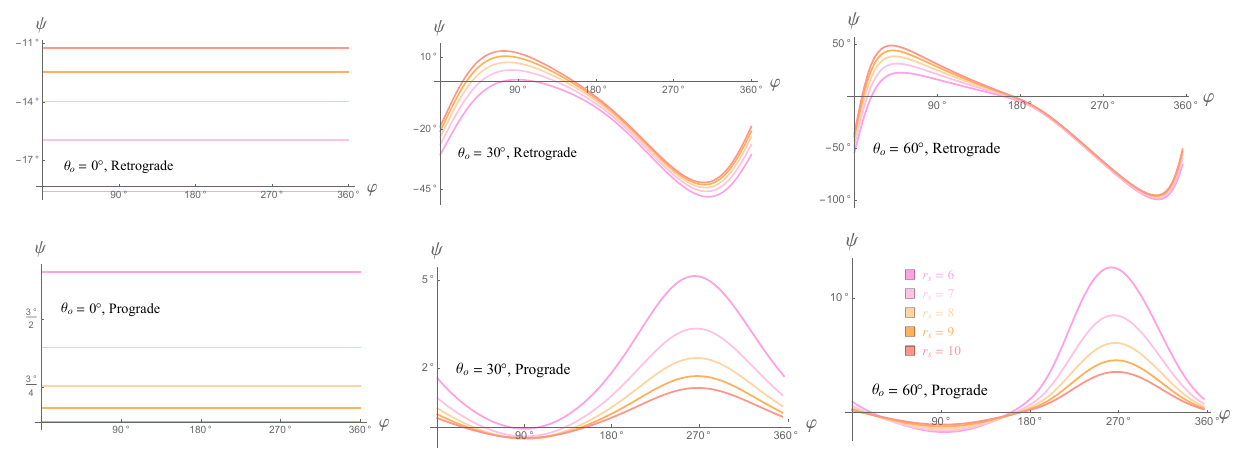}
\caption{The variation of $\psi$ with respect to $\varphi$ for primary images.}
\label{psid}
\end{figure}

On the other side, note that the electric vector position angle (EVPA) is usually defined as
\bea
\text{EVPA}\equiv\frac{1}{2}\arctan \frac{\mathcal{E}_\beta}{\mathcal{E}_\alpha}=\frac{1}{2}\arctan \frac{\beta\kappa_1-\alpha\kappa_2}{\beta \kappa_2+\alpha \kappa_1}
\eea
which is related to our new angle $\psi$ by
\bea
2\text{EVPA}=\varphi+\psi+\frac{1-\text{sign}(\kappa_1)}{2}\pi,
\eea
with $\varphi=\arctan\frac{\beta}{\alpha}$. In this work, we would like to pay more attention to the deviation of the polarization direction from $\varphi$, so we use the angle $\psi$ instead of the EVPA. In addition, there exists an axial symmetry for the image when $\theta_o=0$ and $\psi$ can reflect this property more obviously; see the first column in Fig. \ref{psid}.

Considering the intensities of the secondary images are much lower than those of the primary pictures, we show the results of $\psi$ for the primary images in Fig. \ref{psid}. From Fig. \ref{psid}, we find that for both the prograde and retrograde orbits, $|\psi|_{\text{max}}$ becomes larger as $\theta_o\in [0, \pi/2]$ increases and the orbit radius $r_s\ge 6$ decreases. In particular, when $\theta_o=0$, that is, the observer is at the north or south pole, $\psi=0$ is always true. However, compared with the prograde orbits, $|\psi|_{\text{max}}$ of the retrograde orbits at the same radius is much larger. The polarization structures are significantly different for the prograde and retrograde orbits. In addition, as a warm-up, in this work, we only consider the Schwarzschild black hole, which has no spin. We cannot directly compare the above images with the polarization images taken by EHT. Nevertheless, we may learn some qualitative lesson by comparing two images. From the real polarized image of the supermassive black hole in M87, one can see that the most apparent feature is the shape of the polarization ring, which takes in a closed whorl pattern. As far as the results are concerned, we can see that our models' polarization structures of retrograde orbits are quite similar to those of the polarization image taken by EHT. This could suggest that  the flow of charged particles should be retrograde, and the observational angle should be tiny.

\section{Conclusion}\label{sec5}
In this paper, we have derived two formulas in the geometric optics approximation to describe the direction and the luminosity of polarized synchrotron radiations when a magnetic field accelerates charged particles in curved spacetime.
We applied our formulas to study the polarizations of synchrotron radiations in a Schwarzschild black hole with a vertical uniform magnetic field. We focused on the case that the sources travel
in retrograde or prograde circular orbits on the equatorial plane. We showed the primary and secondary images and polarizations for various inclinations of the observers, as seen in Fig. \ref{diskp} and Fig. \ref{disks}. In particular, we introduced a new angle $\psi$ to quantitatively describe the deviations of the polarization vector from the radial vector at $(\rho, \varphi)$ on the celestial coordinate plane, as shown in Fig. \ref{psid}. Some interesting behaviors of the polarizations were also discussed. Even though our model is based on the synchrotron radiations from the electrons being accelerated by a magnetic field, our results could provide a basis for considering more physical effects theoretically. For example, the charged fluids can be seen as a large number of electrons involving interactions; thus, comparing our results with the results of the fluid, one can distinguish the effects of other physical interactions on the polarization.

In addition, our model can  have important applications in astrophysics. Firstly, our formulas are covariant, making it convenient to be applied to a curved spacetime containing a black hole. Secondly, our model can be seen as a good approximation to describe the polarized images of charged particles in low-density accretion disks. Moreover, our method is suitable for studying the synchrotron radiations of electrons in the magnetosphere, such as the jets. At last, our formulas can be further used to investigate different polarization modes, such as linear and circular polarization. On the other hand, our present results for retrograde charged sources for small observational angles show the shape of the polarization ring takes in a closed whorl, which is consistent with the observed polarized image of the M87 black hole. According to our theoretical results, analyzing the shape of the polarization ring can tell us whether the flow of charged particles is retrograde or prograde, as the first step for further studies.

We conclude this paper with some outlooks. First of all, the environment outside a black hole is always messy. In a realistic rotational spacetime containing a black hole, many complicated ingredients, such as dark matter, jets, halo, etc., should be considered. To have a good insight into the polarized images from EHT, one may have to appeal for an effective metric to describe real spacetime. In this case, the effective metric could become very complicated, and the so-called Penrose-Walker constant does not exist. Nevertheless, our discussion still works well in this situation. Secondly, the equations \eqref{inten} and \eqref{ffeq} may play a fundamental role in studying the radiations of accelerating charged particles in an arbitrary orbit. Based on our work, one could see different features for reflected, bound, and plunging orbits from the polarization information in the images. For example, the radiations from the particles in the plunging orbits may come to us without interacting with the plasma. As a result, they bring "clean" information on the spacetime geometry and the magnetic field configuration to us.

\section*{Acknowledgments}
The work is partly supported by NSFC Grant No. 12205013, 11735001, 11775022, and 11873044. MG is also endorsed by "the Fundamental Research Funds for the Central Universities" with Grant No. 2021NTST13.

%\begin{thebibliography}{10}

%\end{thebibliography}

\end{document}